\documentclass[12pt]{article}

\usepackage{geometry}
\usepackage{cite}
\usepackage{paralist}
\usepackage{graphicx}
\usepackage{subfigure}
\usepackage{amssymb}
\usepackage{amsmath}
\allowdisplaybreaks[1]

\usepackage{amsthm}

\makeatletter
\def\@endtheorem{\endtrivlist}
\makeatother

\newcounter{rrule}
\newenvironment{rrule}{\refstepcounter{rrule}\par\smallskip\noindent
\textbf{(R\arabic{rrule})}\quad}{\par} 

\newcounter{Rule}
\newenvironment{Rule}{\refstepcounter{Rule}\par\smallskip\noindent
\textbf{(\arabic{Rule})}\quad}{} 
\newcommand{\currentrule}{\arabic{Rule}}

\begin{document}

\title{Faster parameterized algorithm for pumpkin vertex deletion set}
\author{Dekel Tsur%
\thanks{Ben-Gurion University of the Negev.
Email: \texttt{dekelts@cs.bgu.ac.il}}}
\date{}
\maketitle

\begin{abstract}
A directed graph $G$ is called a \emph{pumpkin} if $G$ is a union of
induced paths with a common start vertex $s$ and a common end vertex $t$,
and the internal vertices of every two paths are disjoint.
We give an algorithm that given a directed graph $G$ and an integer $k$,
decides whether a pumpkin can be obtained from $G$ by deleting at most $k$
vertices.
The algorithm runs in $O^*(2^k)$ time.
\end{abstract}


\section{Introduction}
A directed graph $G$ is called a \emph{pumpkin} if $G$ is a union of
induced paths with a common start vertex $s$ and a common end vertex $t$,
and the internal vertices of every two paths are disjoint.
The vertices $s$ and $t$ are called the \emph{source} and \emph{sink} of the
pumpkin.
In the \textsc{Pumpkin Vertex Deletion Set} problem (PVDS), the input is
a directed graph $G$ and an integer $k$, and the goal is to decide whether
there is a set of vertices $S$ of size at most $k$ such that the graph obtained
from $G$ by deleting the vertices of $S$ is a pumpkin.
The PVDS problem is NP-hard~\cite{mnich2017polynomial}.
Agrawal et al.~\cite{agrawal2018parameterised} gave an $O^*(2.562^k)$-time
algorithm for PVDS\@.
Polynomial kernels for this problem were given
in~\cite{mnich2017polynomial,agrawal2018kernels}.

In this paper, we give an $O^*(2^k)$-time algorithm for PVDS\@.

\section{Preliminaries}
For a vertex $v$ in a directed graph $G$, $N^-_G(v)$ and $N^+_G(v)$ denote the
set of in-neighbors and out-neighbors of $v$.
Additionally, $d^-_G(v) = |N^-_G(v)|$ and $d^+_G(v) = |N^+_G(v)|$.
We will sometimes omit the subscript $G$ if the graph $G$ is clear from the
context.

For set of vertices $S$, $G-S$ is the graph obtained from $G$ by deleting
the vertices of $S$ (and incident edges).

\section{The algorithm}

In the \textsc{Restricted Pumpkin Vertex Deletion Set} problem (RPVDS),
the input is a directed graph $G$, an integer $k$ and vertices $s,t$.
The goal is to decide whether there is a set of vertices $S$ of size
at most $k$ such that the graph obtained from $G$ by deleting the vertices
of $S$ is a pumpkin with source $s$ and sink $t$.
An $O^*(c^k)$-time algorithm for RPVDS implies an $O^*(c^k)$-time algorithm
for PVDS: Given an input $(G,k)$ to PVDS, go over all possible ways to
choose $s$ and $t$, and for each choice run the RPVDS algorithm on $(G,k,s,t)$.
The time complexity is $O^*(n^2 \cdot c^k) = O^*(c^k)$.
In the sequel, we describe an $O^*(2^k)$-time algorithm for RPVDS.

Our algorithm is a branching algorithm (cf.~\cite{cygan2015parameterized}).
Given an instance $(G,k,s,t)$, the algorithm applies the first applicable reduction
rule from the rules below, and if no reduction rule is applicable, the algorithm
applies the first applicable branching rule from the rules below.

The algorithm uses the following reduction rules from
Agrawal et al.~\cite{agrawal2018parameterised} (we present the rules in slightly simplified
form).

\begin{rrule}
If $k < 0 $ return `no'.
\end{rrule}

\begin{rrule}
If $k = 0$ and $G$ is not a pumpkin with source $s$ and sink $t$ return `no'.
\end{rrule}

\begin{rrule}
If $G$ is a pumpkin with source $s$ and sink $t$ return `yes'.
\end{rrule}

\begin{rrule}
If there is a vertex $v$ that is not reachable from $s$,
delete $v$ from $G$ and decrease $k$ by 1.
\end{rrule}

\begin{rrule}
If there is a vertex $v\neq t$ such that $t$ is not reachable from $v$,
delete $v$ from $G$ and decrease $k$ by 1.
\end{rrule}

\begin{rrule}
If there is a vertex $v$ such that $s\in N^+(v)$,
delete $v$ from $G$ and decrease $k$ by 1.
\end{rrule}

\begin{rrule}
If there is a vertex $v$ such that $t \in N^-(v)$,
delete $v$ from $G$ and decrease $k$ by 1.
\end{rrule}


\begin{rrule}
If $t$ is not reachable from $s$, return `no'.
\end{rrule}

We now describe the branching rules of the algorithm.
When we say that the algorithm branches on sets $S_1,\ldots,S_p$, we mean
that the algorithm is called recursively on the instances
$(G-S_1,k-|S_1|,s,t),\ldots,(G-S_p,k-|S_p|,s,t)$.
For each Rule~($i$) below, except Rule~(\ref{rule:double}), there is also
a symmetric rule, denoted Rule~($i'$), in which the roles of in-neighbors
and out-neighbors are reversed.
For example, Rule~(\ref{rule:t}$'$) is:
If there is a vertex $v\neq s,t$ such that $ d^-(v) \geq 2$ and $s \in N^-(v)$,
branch on $\{v\}$ and $N^-(v) \setminus \{s\}$.
The order in which the branching rules are considered is (1), (2), ($2'$), (3),
($3'$) and so on.
\begin{Rule}
If there are vertices $u$ and $v$ such that $(u,v)\in E(G)$ and
$(v,u)\in E(G)$, branch on $\{u\}$ and $\{v\}$.
\label{rule:double}
\end{Rule}

The correctness of Rule~(\currentrule) is trivial.
The branching vector of Rule~(\currentrule) is $(1,1)$.

\begin{Rule}
If there is a vertex $v\neq s,t$ such that $ d^+(v) \geq 2$ and $t \in N^+(v)$,
branch on $\{v\}$ and $N^+(v) \setminus \{t\}$.
\label{rule:t}
\end{Rule}

To prove the correctness of Rule~(\currentrule), note that if $S$ is
a solution to the instance $(G,k,s,t)$ and $v \notin S$ then
$N^{+}(v) \setminus\{t\} \subseteq S$,
otherwise $d^+_{G-S}(v) \geq 2$ contradicting the assumption that $S$ is a
solution.
The branching vector of Rule~(\currentrule) is at least $(1,1)$.


\begin{Rule}
If there is a vertex $v\neq s,t$ such that $d^+(v) \geq 4$,
arbitrarily choose distinct $w_1,w_2,w_3,w_4 \in N^+(v)$.
Branch on $\{v\}$ and on $\{w_1,w_2,w_3,w_4\}\setminus\{w_i\}$ for every $i\leq 4$.
\label{rule:deg-4}
\end{Rule}

The correctness of Rule~(\currentrule) is obvious: If $S$ is
a solution to the instance $(G,k,s,t)$ and $v \notin S$ then $d^+_{G-S}(v) = 1$.
Therefore, $S$ contains at least $3$ vertices of $\{w_1,w_2,w_3,w_4\}$.
The branching vector of Rule~(\currentrule) is $(1,3,3,3,3)$.

\begin{Rule}
If there is a vertex $v\neq s,t$ such that $d^+(v) \geq 2$
and there is a vertex $w\in N^+(v)$ for which $d^-(w) = 1$,
branch on $\{w\}$ and $N^+(v) \setminus \{w\}$.
\label{rule:degree-1-neighbor}
\end{Rule}

We now prove the correctness of Rule~(\currentrule).
Note that $w \neq s$ otherwise Rule~(R6) can be applied on $v$, a contradiction.
Let $S$ be a solution to $(G,k,s,t)$.
If $w \notin S$ then we also have $v \notin S$
(otherwise $d^-_{G-S}(w) = 0$ and since $w \neq s$ this is a contradiction to
the assumption that $S$ is a solution).
Therefore, $S$ contains all the vertices in $N^+(v) \setminus \{w\}$
(otherwise $d^+_{G-S}(v) \geq 2$).
The branching vector of Rule~(\currentrule) is at least $(1,1)$.

\begin{Rule}
If there is a vertex $v\neq s,t$ such that there are $w_1,w_2 \in N^+(v)$
for which $(w_1,w_2)\in E(G)$ and $N^+(w_1) =\{w_2\}$,
branch on $\{v\}$ and $\{w_1\}$.
\label{rule:edge}
\end{Rule}

To prove the correctness of Rule~(\currentrule),
suppose that $S$ is a solution to $(G,k,s,t)$ and $v \notin S$.
We claim that $w_1 \in S$.
Suppose conversely that $w_1 \notin S$.
Note that $w_1 \neq t$ otherwise Rule~(\ref{rule:t}) can be applied
on $v$, a contradiction.
Therefore, $w_2 \notin S$ (otherwise $d^+_{G-S}(w_1) = 0$
and since $w_1 \neq t$ we obtain a contradiction).
Since $w_1,w_2 \in N^+(v)$, it follows that $d^+_{G-S}(v) \geq 2$,
a contradiction.
The branching vector of Rule~(\currentrule) is $(1,1)$.

\begin{Rule}
If there is a vertex $v\neq s,t$ such that $d^+(v) = 3$ and there are at least
two vertices $w_1,w_2$ in $N^+(v)$ for which
$N^-(w_i) \setminus \{v\} \not\subseteq \cup N^+(v)$ for $i = 1,2$,
denote by $w_3$ the third vertex in $N^+(v)$.
Branch on $\{v\}$,
$\{w_2,w_3\} \cup (N^-(w_1) \setminus \{v\})$,
$\{w_1,w_3\} \cup (N^-(w_2) \setminus \{v\})$ and $\{w_1,w_2\}$.
\label{rule:deg-3-1}
\end{Rule}

If $S$ is a solution to $(G,k,s,t)$ and $v\notin S$ then $S\cap \{w_1,w_2,w_3\}$
is either $\{w_2,w_3\}$, $\{w_1,w_3\}$, or $\{w_1,w_2\}$
(otherwise $d^+_{G-S}(v) \neq 1$).
In the first case we have that $N^-(w_1) \setminus \{v\} \subseteq S$
(otherwise $d^-_{G-S}(w_1) \geq 2$)
and in the second case $N^-(w_2) \setminus \{v\} \subseteq S$.
The branching vector of Rule~(\currentrule) is at least $(1,3,3,2)$.

\begin{Rule}
If there is a vertex $v\neq s,t$ such that $d^+(v) = 3$,
denote $N^+(v) = \{w_1, w_2, w_3\}$ where $N^-(w_2) = \{v, w_1\}$
(we will show below that there is always such numbering of the vertices in
$N^+(v)$).
Branch on $\{w_2\}$, $\{w_1,w_3\}$, and
$\{v\} \cup (N^+(w_1) \setminus \{w_2\})$.
\label{rule:deg-3-2}
\end{Rule}

Since Rule~(\ref{rule:deg-3-1}) cannot be applied, there are
two vertices $w_2,w_3$ in $N^+(v)$ for which 
$N^-(w_i) \setminus \{v\} \subseteq N^+(v)$ for $i = 2,3$.
Let $w_1$ be the third vertex in $N^+(v)$.
Since there is at most one edge whose endpoints are $w_2$ and $w_3$,
without loss of generality $(w_2,w_3) \notin E(G)$.
Since Rule~(\ref{rule:degree-1-neighbor}) cannot be applied,
$d^-(w_2) \geq 2$.
It follows that $N^-(w_2) = \{v, w_1\}$.

Let $S$ be a solution to $(G,k,s,t)$ and suppose that $w_2 \notin S$.
Since $d^-_{G-S}(w_2) = 1$, either $v \notin S$ or $w_1 \notin S$.
In the former case $w_1,w_3 \in S$ (otherwise $d^+_{G-S}(v) \geq 2$).
In the latter case $v \in S$ (otherwise $d^+_{G-S}(v) \geq 2$)
and $N^+(w_1) \setminus \{w_2\} \subseteq S$
(otherwise $d^+_{G-S}(w_1) \geq 2$).
Note that $N^+(w_1) \setminus \{w_2\} \neq \emptyset$ otherwise
Rule~(\ref{rule:edge}) can be applied, a contradiction.
The branching vector of Rule~(\currentrule) is at least $(1,2,2)$.

Note that if Rules (R1)--(R8), (1)--(\currentrule), and
(2$'$)--(\currentrule$'$) cannot be applied then
$d^-(v) \in \{1,2\}$ and $d^+(v) \in \{1,2\}$ for every vertex $v \neq s,t$.
\begin{Rule}
If there is a vertex $v\neq s,t$ such that $d^+(v) = 2$,
let $v$ be such vertex whose distance from $s$ is minimal.
Denote $N^+(v) = \{w_1, w_2\}$ such that $(w_1,w_2) \notin E(G)$.
We have that $d^-(w_2) = 2$
(otherwise Rule~(\ref{rule:degree-1-neighbor}) can be applied on $v$),
and let $x$ be the single vertex in $N^-(w_2) \setminus \{v\}$.
Moreover, $d^+(x) = 2$ (otherwise Rule~(\ref{rule:degree-1-neighbor}$'$) can
be applied on $w_2$).
Let $y$ be the single vertex in $N^+(x) \setminus \{w_2\}$.
branch on $\{w_2\}$, $\{w_1, x\}$, and $\{v, y\}$.
\label{rule:deg-2}
\end{Rule}

Note that $x \neq s,t$ (due to Rule~(\ref{rule:t}$'$) and Rule~(R7)).
Additionally, $x \neq w_1$ due to the assumption that $(w_1,w_2) \notin E(G)$.
If $S$ is a solution to $(G,k,s,t)$ and $w_2 \notin S$ then
either $v \notin S$ or $x \notin S$.
In the former case $w_1 \in S$ (otherwise $d^+_{G-S}(v) \geq 2$)
and $x \in S$ (otherwise $d^-_{G-S}(w_2) \geq 2$).
In the latter case $v \in S$ (otherwise $d^-_{G-S}(w_2) \geq 2$) and
$y \in S$ (otherwise $d^+_{G-S}(x) \geq 2$).

To get the desired branching vector, we show that $y \neq v$.
Suppose conversely that $y = v$.
Therefore, $d^-(v) = 2$ (otherwise Rule~(\ref{rule:degree-1-neighbor}) can be
applied on $x$).
Let $z$ be the single vertex in $N^-(v) \setminus \{x\}$.
We have that $d^+(z) = 2$ (otherwise Rule~(\ref{rule:degree-1-neighbor}$'$) can
be applied on $v$).
By definition, $N^-(v) = \{x,z\}$.
Therefore, at least one of $x$ and $z$ have smaller distance from $s$ than $v$.
Since $d^+(x) = d^+(z) = 2$, we obtain a contradiction to the definition of $v$.
Therefore, $y \neq v$,
and thus the branching vector of Rule~(\currentrule) is $(1,2,2)$.

The branching vectors of the branching rules of the algorithm are
$(1,1)$, $(1,2,2)$, $(1,3,3,2)$, and $(1,3,3,3,3)$ (in the worst cases).
All these vectors have branching number $2$.
Therefore, the running time of the algorithm is $O^*(2^k)$.

\bibliographystyle{plain}
\bibliography{pumpkin}

\begin{thebibliography}{1}

\bibitem{agrawal2018kernels}
Akanksha Agrawal, Saket Saurabh, Roohani Sharma, and Meirav Zehavi.
\newblock Kernels for deletion to classes of acyclic digraphs.
\newblock {\em Journal of Computer and System Sciences}, 92:9--21, 2018.

\bibitem{agrawal2018parameterised}
Akanksha Agrawal, Saket Saurabh, Roohani Sharma, and Meirav Zehavi.
\newblock Parameterised algorithms for deletion to classes of {DAGs}.
\newblock {\em Theory of Computing Systems}, pages 1--30, 2018.

\bibitem{cygan2015parameterized}
Marek Cygan, Fedor~V Fomin, {\L}ukasz Kowalik, Daniel Lokshtanov, D{\'a}niel
  Marx, Marcin Pilipczuk, Micha{\l} Pilipczuk, and Saket Saurabh.
\newblock {\em Parameterized algorithms}.
\newblock Springer, 2015.

\bibitem{mnich2017polynomial}
Matthias Mnich and Erik~Jan van Leeuwen.
\newblock Polynomial kernels for deletion to classes of acyclic digraphs.
\newblock {\em Discrete Optimization}, 25:48--76, 2017.

\end{thebibliography}

\end{document}